\documentclass[onecolumn,showkeys,showpacs,notitlepage,superscriptaddress]{revtex4-1}
\usepackage{amsmath}
\usepackage{graphicx}
\usepackage{hyperref}
\usepackage{color}

\newcommand{\beginsupplement}{%
\setcounter{figure}{0}
\renewcommand{\thefigure}{S\arabic{figure}}%
}

\begin{document}
\title{Statistical physics methods provide the exact solution to a long-standing problem of genetics}
\author{Areejit Samal}
\affiliation{Laboratoire de Physique Th\'eorique et Mod\`eles Statistiques (LPTMS), CNRS and Univ Paris-Sud, UMR 8626, F-91405 Orsay, France}
\affiliation{Max Planck Institute for Mathematics in the Sciences, 04103 Leipzig, Germany}
\affiliation{The Abdus Salam International Centre for Theoretical Physics, Trieste 34151, Italy}
\author{Olivier C. Martin}
\email{olivier.martin@moulon.inra.fr}
\affiliation{INRA, UMR 0320 / UMR 8120 G\'en\'etique Quantitative et Evolution -- Le Moulon,
F-91190 Gif-sur-Yvette, France}
\begin{abstract}
Analytic and computational methods developed within statistical physics have found applications in numerous disciplines. In this letter, we use such methods to solve a long-standing problem in {\it statistical genetics}. The problem, posed by Haldane and Waddington [J.B.S. Haldane and C.H. Waddington, Genetics {\bf 16}, 357-374 (1931)], concerns so-called recombinant inbred lines (RILs) produced by repeated inbreeding. Haldane and Waddington derived the probabilities of RILs when considering 2 and 3 genes but the case of 4 or more genes has remained elusive. Our solution uses two probabilistic frameworks relatively unknown outside of physics: Glauber's formula and self-consistent equations of the Schwinger-Dyson type. Surprisingly, this combination of statistical formalisms unveils the {\it exact} probabilities of RILs for {\it any} number of genes. Extensions of the framework may have applications in population genetics and beyond.
\end{abstract}
\maketitle
\begin{figure}
\includegraphics[width=.5\columnwidth]{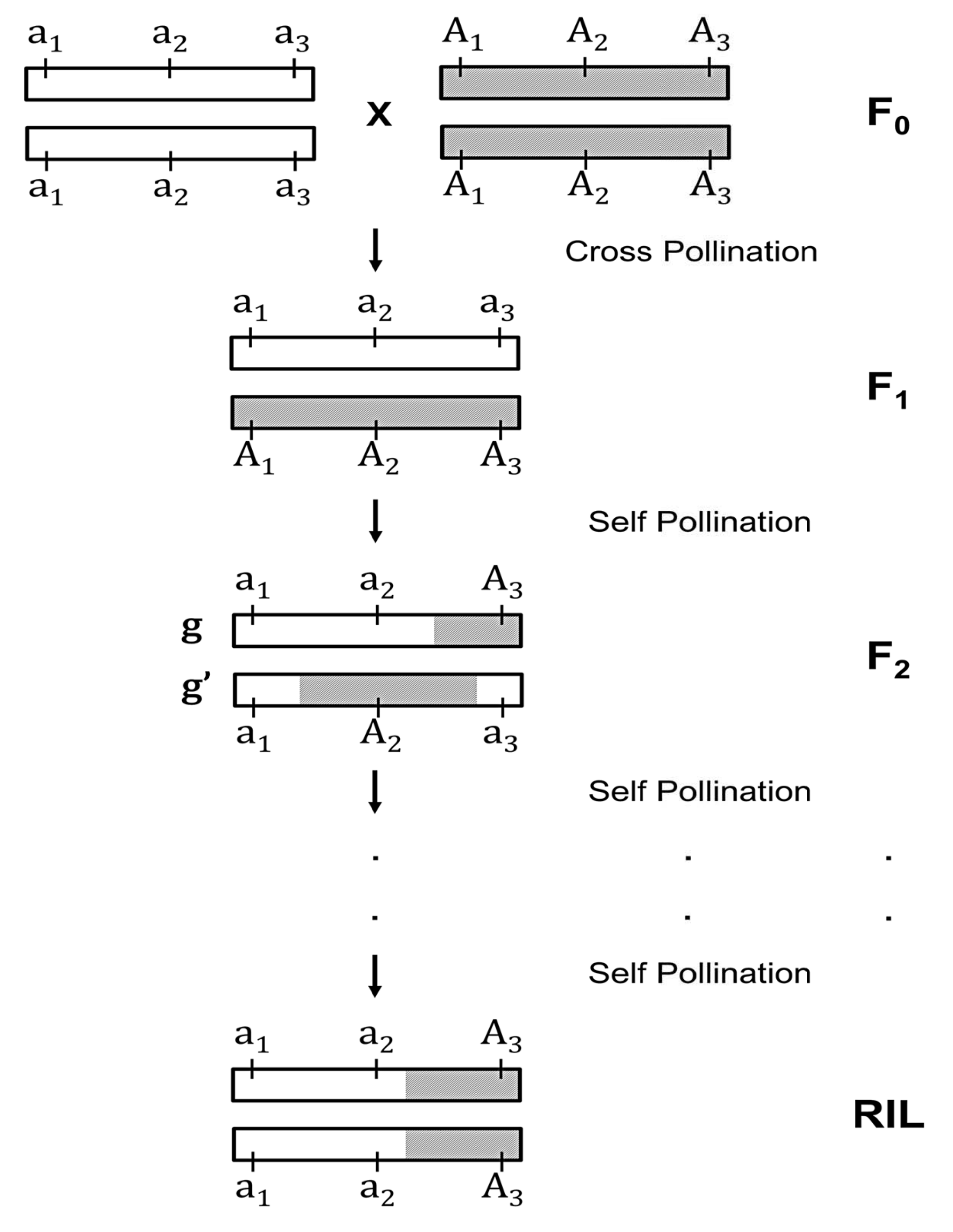}
\caption{{\bf Production of one Recombinant Inbred Line.}
A chromosome pair is followed in each plant. A new generation results from 2 gametes that may mix genetic content as shown via $g$ and $g'$. Tracking of allelic types (``a'' and ``A'') is displayed at 3 positions until no further change is possible.}
\label{RILfigure}
\end{figure}

Statistical physics methods have fertilized numerous disciplines including complex networks~\cite{Albert_2002}, theoretical computer science~\cite{Mezard_2002} and Bayesian statistical inference~\cite{Toussaint_2002}. They have also led to novel results in population genetics~\cite{StatPhysToGenetics}. Here we use those methods to tackle an old problem of genetics involving recombinant inbred lines (RILs). A RIL is  produced via repeated inbreeding of animals or plants until all genetic variability has been removed (see Fig.~\ref{RILfigure}). The individuals produced in this way constitute a stable and permanently shareable genetic resource that is particularly useful for the identification of genes contributing to traits of interest~\cite{Crow_2007}. These properties explain why production and exploitation of large populations of RILs have become major endeavors in the search for genetic determinants of diseases in mammals~\cite{CTC} and of agricultural traits in crops~\cite{Buckler_2009}.

In this letter, we consider plant RILs that are produced using {\it Single Seed Descent} (SSD) which is an extreme form of inbreeding. One starts with two founding parents that are ``homozygous'' everywhere, \emph{i.e.}, for each pair of chromosomes, the two associated alleles are identical. This situation is schematically represented in Fig.~\ref{RILfigure} using the generation label $F_0$ and by displaying a single \emph{pair} of chromosomes for each plant. The two parents being genetically different, their chromosomal contents are shown using different shadings. These two parents are then cross pollinated: one parent produces a female gamete while the other parent produces a male gamete. The fusion of the two gametes will lead to the \emph{single} F1 plant at the next generation. Consider going now from generation $F_1$ to generation $F_2$. Cross pollination is replaced by self pollination: the single F1 plant produces \emph{both} the female gamete ($g$) and the male gamete ($g'$). This capability arises in almost all plants of agricultural interest. A subtlety now arises as shown in Fig.~\ref{RILfigure}: a gamete can form a mosaic of the two chromosomes from which it is built. This phenomenon follows from the formation of ``crossovers'' between the two chromosomes during gamete formation. It can occur at all generations but in the case of going from $F_0$ to $F_1$ it simply has no visible effects. The process of producing a RIL is based on iterating the step when going from $F_1$ to $F_2$: \emph{self pollination} of a single $F_n$ plant is used to produce a seed which develops into the single $F_{n+1}$ plant, thus the term {\it Single Seed Descent}. Note that once a chromosomal region has become homozygous, (in the figure this corresponds to having locally the same shading for the two chromosomes), it stays so. (If a region is not homozygous, one says it is heterozygous.) Thus, because of chance, after enough generations, the plant becomes homozygous everywhere. The chromosomes of the resulting RIL are mosaics of the two parental chromosomes at $F_0$. Given many such RILs (cf.~\cite{SM-RIL}), statistical inference can be used to identify the chromosomal regions responsible for parental differences in traits of interest~\cite{Buckler_2009}.

Experimentally, one often determines a plant's genetic content at discrete positions or ``loci''; we assign these an index $i$ ranging from 1 to $L$ (from left to right along the chromosome). Denote by ``a'' the allelic type (white) of the first parent and by ``A'' that (shaded) of the second parent. Then the {\it genotype} of parent ``a'' is $\left( a_1/a_1,a_2/a_2,\ldots,a_L/a_L \right)$ and that of parent ``A'' is $\left( A_1/A_1,A_2/A_2,\ldots,A_L/A_L \right)$ where the $-_i/-_i$ notation provides the allelic type on the two chromosomes for ``locus'' $i$. Fig.~\ref{RILfigure} illustrates a case with $L=3$ for which both gametes have crossovers when going from F1 to F2. The allelic type of the first gamete changes when going from locus 2 to locus 3; one says that the interval (2,3) is ``recombinant'' or that there has been a recombination event between the two loci in that gamete.

\begin{figure*}
\includegraphics[width=16cm]{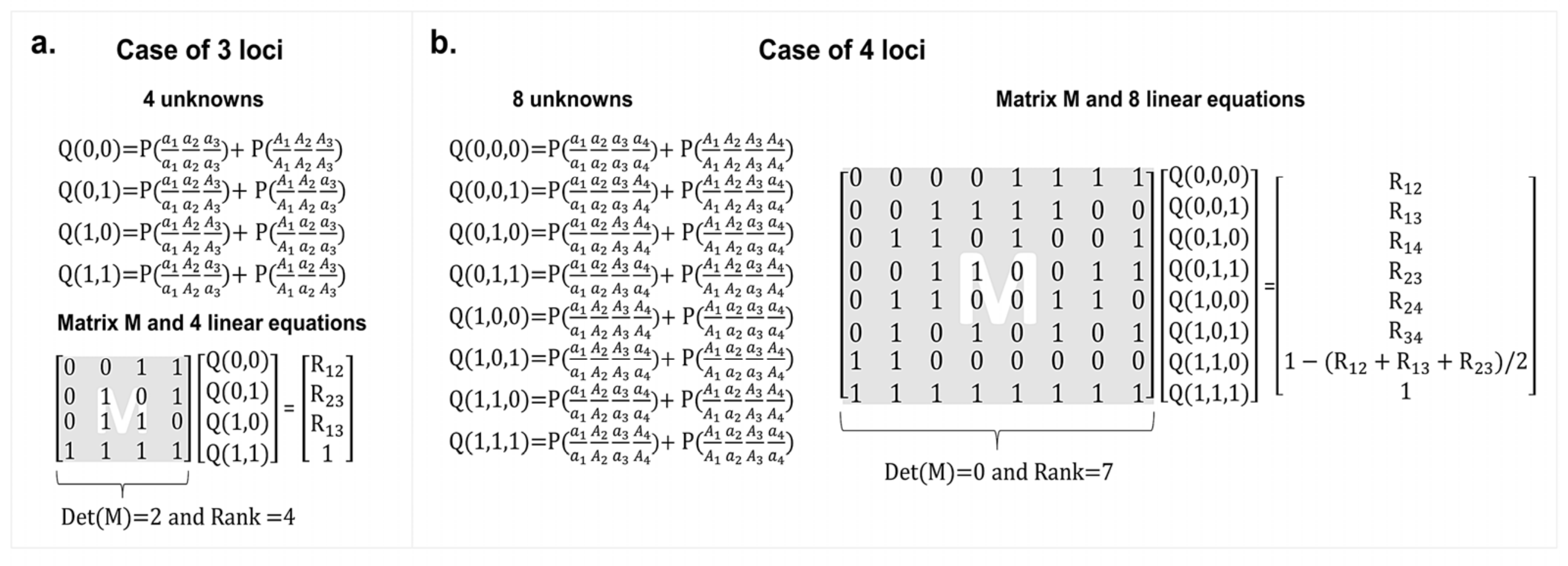}
\caption{{\bf The 2- and 3-locus RIL probabilities do not completely specify the 4-locus RIL probabilities.} (a) The matrix of the linear equations relating the 3-locus probabilities to 2-locus probabilities ({\it via} the $R_{i,j}$'s) has Rank 4. (b) A matrix giving linear equations relating the 4-locus probabilities to 2- and 3-locus probabilities always has Rank at most 7.}
\label{MissingQuantity}
\end{figure*}

In 1918 Robbins~\cite{Robbins_1918} determined the probabilities of 2-locus RIL genotypes produced using SSD. Then, in 1931, Haldane and Waddington~\cite{Haldane_Waddington_1931} simplified that derivation. Based on meiotic recombination rates independent of allelic content and of sex, they provided the celebrated Haldane-Waddington formula~\cite{Haldane_Waddington_1931} giving the ``RIL recombination rate'' between 2 loci $i$ and $j$, \emph{i.e.}, $R_{i,j} = 2 r_{i,j} / (1+2 r_{i,j})$ where $r_{i,j}$ is the $(i,j)$ recombination rate {\it per meiosis}.  Probabilities of all 2-locus RIL genotypes are then directly obtained using the definition of the RIL recombination rate: $R_{i,j} = P( a_i/a_i, A_j/A_j ) + P( A_i/A_i, a_j/a_j )$ which is the probability that the alleles will be recombined after enough inbreeding. By symmetry, $P(  a_i/a_i, A_j/A_j ) = P(  A_i/A_i, a_j/a_j ) = R_{i,j}/2$ and $P(  a_i/a_i, a_j/a_j )= P(  A_i/A_i, A_j/A_j ) = (1-R_{i,j})/2$ (\cite{SM-TwoSpin}).

In 1931 Haldane and Waddington~\cite{Haldane_Waddington_1931} also showed that the 2-locus RIL probabilities determine the ones for 3 loci. Over time, the results for 2 and 3 loci have been refined or extended to other kinds of crosses~\cite{Refinement}, but the case of 4 or more loci has proved to be inextricable. This fact appears as particularly puzzling since going from 2 to 3 loci is very simple and involves just standard algebra (see Fig. \ref{MissingQuantity}a and \cite{SM-contrast}). The point is that 2- and 3-locus RIL probabilities do {\it not} determine the 4-locus probabilities (see Fig. \ref{MissingQuantity}b and \cite{SM-contrast}). Finding and exploiting this missing information has prevented researchers from extending the Haldane-Waddington result for over 80 years. In this letter, we provide a solution to this challenge, deriving exact analytic formulas for the probabilities of RIL genotypes having {\it any} number of loci. The breakthrough is based on using two probabilistic frameworks borrowed from physics: the Schwinger-Dyson equations~\cite{Dyson_1949,Schwinger_1951} and Glauber's formula~\cite{Glauber_1963}.

Given that a RIL is homozygous at every locus, its genetic content can be specified in terms of a vector $\vec{S}$ of {\it spin} variables $S_i$, $i=1, 2, \ldots L$. Our convention, motivated by~\cite{Slatkin_1972}, is $S_i=1$ if locus $i$ is $a_i/a_i$ and $S_i=-1$ if it is $A_i/A_i$. This notation is particularly convenient for writing the probability of any RIL genotype $\vec{S}$ in terms of averages of spin products. For example, if there is a single locus $i$, the probability that the spin has value $s_i$ is $P(S_i=s_i) = E [ (1 + s_i S_i)/2 ]$ where the average or expectation $E[~]$ is taken over the distribution of the random variable $S_i$. For $L$ loci, the generalization of this formula, due to Glauber~\cite{Glauber_1963}, is
\begin{equation}
\label{GlauberFormula}
\begin{split}
P(\{ S_1=s_1, S_2=s_2, \ldots S_L=s_L \}) = \\
E [ (\frac{1 + s_1 S_1}{2})(\frac{1 + s_2 S_2}{2}) \ldots (\frac{1 + s_L S_L}{2}) ]
\end{split}
\end{equation}
where $E[~]$ is the average over all possible RIL genotypes with their corresponding probabilities. Note that Eq. \ref{GlauberFormula} is exact, the $S_i$ need not be independent. The problem of finding the probabilities of all RIL genotypes is then solved if one can determine the expectation values of all spin products. When expanding the right-hand side of Eq. \ref{GlauberFormula}, expectation values of $k$-allelic products come with a sign equal to the product of the corresponding $s_i$ values. For instance for $L=4$, Eq. \ref{GlauberFormula} leads to
\begin{equation}
\label{GlauberFormula4Loci}
\begin{split}
P(\{ S_1=s_1,S_2=s_2,S_3=s_3,S_4=s_4 \}) = \\
\frac{1}{16} ( 1 + \sum_{i < j}  s_i s_j E [ S_i S_j ] +  s_1 s_2 s_3 s_4 E [ S_1 S_2 S_3 S_4 ] )
\end{split}
\end{equation}
where we have used the fact that the expectation of a product of an odd number of $S_i$'s vanishes because of the global invariance $P(\vec{S}) = P(-\vec{S})$, corresponding to exchanging all ``a''s and ``A''s in RIL genotypes.

To explain our approach, we begin by solving the 4-locus case ($L=4$). Eq. \ref{GlauberFormula4Loci} shows that we need the expectations of 2- and 4-spin products. The 2-spin products are given by $E [ S_i S_j ] = 1 - 2 R_{i,j}$ \cite{SM-TwoSpin} so the only unknown is the 4-spin product $E [ S_1 S_2 S_3 S_4 ]$ in direct correspondence with the situation described in Fig. \ref{MissingQuantity}b. Our strategy to compute $E [ S_1 S_2 S_3 S_4 ]$ is based on classifying the ways of going from the first generation of children (F1) all the way to the RIL according to the genotype arising at the second generation of children (F2) (Fig. \ref{RILfigure}). Performing this classification leads to
\begin{equation}
\label{FourSpinProduct}
E [  S_1 S_2 S_3 S_4 ] = \sum_g \sum_{g'} P(g)P(g') E_{g,g'} [ S_1 S_2 S_3 S_4 ]
\end{equation}
where the sum is over all F2 genotypes (each specified by the genotypes of its female ($g$) and male ($g'$) gametes), $P(g)$ is the probability of producing a gamete of genotype $g$ when going from F1 to F2, and $E_{g,g'} [ S_1 S_2 S_3 S_4 ]$ is the expectation of the 4-spin product {\it when starting the inbreeding with an F2 individual of genotype $(g,g')$}. Now the key point is that $E_{g,g'} [ S_1 S_2 S_3 S_4 ]$ is equal to $E [ S_1' S_2' S_3' S_4' ]$ when starting with the F1 if one uses the following {\it substitution} rules for the $S_i'$. First, if locus $i$ is homozygous in $G=(g,g')$ and has value $s_i$, then all descendants of $G$ also have that value, so replace $S_i'$ by $s_i$. Second, if locus $i$ is heterozygous in $G$ and is of the type $a_i/A_i$, the situation is the same as at F1, so replace $S_i'$ by $S_i$. Finally, if locus $i$ is heterozygous in $G$ and is of the type $A_i/a_i$, {\it i.e.}, it is reversed compared to the F1, replace $S_i'$ by $-S_i$. These simple rules provide the way to relate expectations starting with an F2 genotype to expectations starting with the F1 genotype. The self-consistent Eq. \ref{FourSpinProduct} then becomes a Schwinger-Dyson (SD) equation~\cite{Dyson_1949,Schwinger_1951} where the expectation value of the 4-spin product (on the left) is expressed (on the right) in terms of itself and of lower order spin-product averages. By summing the contributions of the $4^4$ different F2 genotypes in Eq. \ref{FourSpinProduct}, we can extract the value for $E [ S_1 S_2 S_3 S_4 ]$ and then our problem is solved, {\it i.e.}, Eq. \ref{GlauberFormula4Loci} provides all 4-locus probabilities.

\begin{figure}
\includegraphics[width=.5\columnwidth]{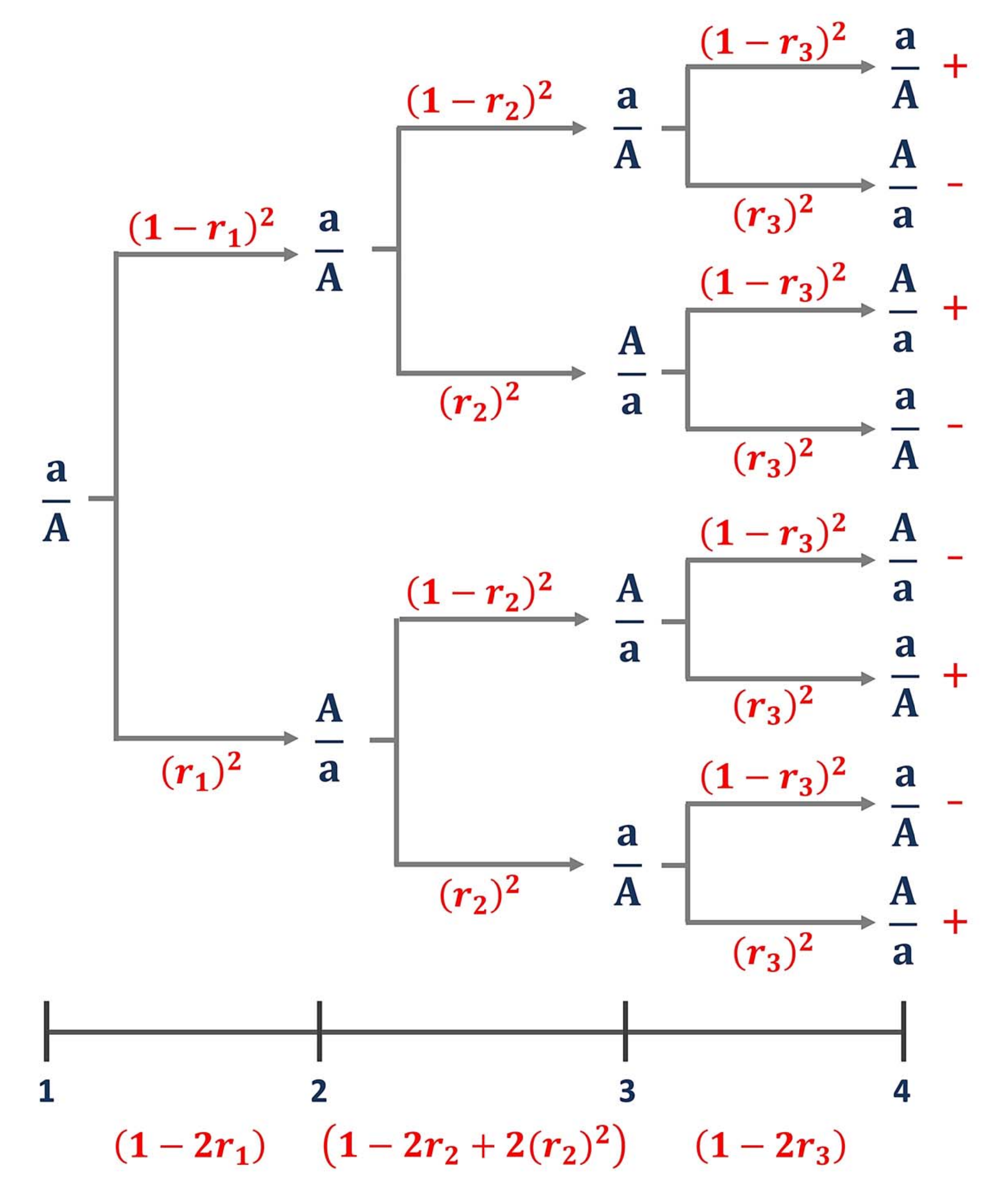}
\caption{{\bf Tree-mapping of F2 genotypes.} F2 genotypes map to paths from root to leaves of trees. The sign of a genotype is given on the right and its weight is the product of factors along the path. Summing over all paths of this tree leads to the factors shown at bottom.}
\label{Treehhhh}
\end{figure}

In Eq. \ref{FourSpinProduct} the sum over all F2 genotypes involves the probabilities $P(g)$. If the crossovers arise independently as in Haldane's no interference model~\cite{Haldane_1919}, then the summation in Eq. \ref{FourSpinProduct} can be performed by hand and very elegantly as follows.  First we regroup the F2 genotypes into classes according to which of their loci are heterozygous. For each class the associated contributions can be summed explicitly by mapping to a tree. To see how this works, consider for instance calculating the factor multiplying $E[ S_1 S_2 S_3 S_4 ]$ on the right-hand side of the SD equation.  This factor is obtained by considering the class of F2 genotypes that are heterozygous  ($h$) at all 4 loci and calculating the sum over those F2 genotypes of the probability ($P(g) P(g')$) times the sign from the substitution rule. Fig. \ref{Treehhhh} represents the mapping of these genotypes, their probabilities and their signs onto a tree for the case where the first locus is of the type $a_1/A_1$. The loci are ordered from left to right and each F2 genotype can be identified with a path from the left-most node to one of the right-most nodes (leaves of the tree).  Due to the assumption of no crossover interference, recombination arises independently in each interval so that the probability of a genotype can be written as a product of factors, one for each interval.  For any specified interval in Fig. \ref{Treehhhh}, the 2 gametes are either both non-recombinant, leading to a factor $(1-r_i)^2$, or both recombinant, leading to a factor $r_i^2$, where $r_i$ is the recombination rate for a single meiosis in the interval $(i,i+1)$. The probability of a F2 genotype is then given by the product of such factors along the path as displayed in Fig. \ref{Treehhhh}, times $1/4$ coming from the probability that the first locus is of the type $a_1/A_1$. Adding the contributions of all genotypes of the tree shown in Fig. \ref{Treehhhh} can be done by recurrence \cite{SM-FourLoci}. Using the fact that the tree rooted at $A_1/a_1$ gives rise to the same calculation as for Fig.~\ref{Treehhhh}, one concludes that the class of heterozygous genotypes on the right-hand side of Eq.~\ref{FourSpinProduct} contributes a total of $(1 - 2 r_1)  ( (1-r_2)^2 + r_2^2 ) (1 - 2 r_3)/2$ times $E[ S_1 S_2 S_3 S_4 ]$.

The other classes can be treated by the same mapping technique. Consider for instance the class of F2 genotypes homozygous at all loci. It is easy to see that it leads to exactly the same result as the class just treated except that $E[ S_1 S_2 S_3 S_4 ]$ is replaced by 1 \cite{SM-FourLoci}. Going on to the classes which are mixed (with both homozygous and heterozygous loci), only those having two adjacent loci homozygous and two adjacent loci heterozygous lead to non-zero contributions \cite{SM-FourLoci}. In those cases, between the second and third locus, there is one and only one recombinant gamete, whereas in the previous calculation in that interval $g$ and $g'$ were both recombinant or both non-recombinant. Thus the previously derived term $( (1-r_2)^2 + r_2^2 )$ has to be replaced by $2 r_2 (1-r_2)$ here (Fig. S2 in \cite{SM-FourLoci}). Collecting the results from all classes of F2 genotypes leads to the 4-locus SD equation:
\begin{widetext}
\begin{equation}
\label{SD4loci}
E[ S_1 S_2 S_3 S_4 ] = \frac{(1 - 2 r_1)  ( (1-r_2)^2 + r_2^2 ) (1 - 2 r_3)}{2} ( E[ S_1 S_2 S_3 S_4 ] + 1) + \frac{(1 - 2 r_1) (2 (1-r_2) r_2 ) (1-2 r_3)}{2} ( E[ S_1 S_2 ] + E[ S_3 S_4 ] )
\end{equation}
\end{widetext}

Although the expectation of the 4-spin product arises on both sides of this equation, extracting this quantity in terms of the averages of 2-spin products is straightforward. In summary, from Eq. \ref{GlauberFormula4Loci}, using Eq. \ref{SD4loci} and the formula $E[ S_i S_j ] = 1 - 2 R_{i,j}$, one obtains the long-searched-for exact analytic expressions for 4-locus RIL genotype probabilities.

The overall framework, including the mapping of F2 genotypes to trees, extends to any number of loci. For 5 loci, no new SD equation is needed since the expectation $E [ S_1 S_2 \ldots S_L ]$ vanishes when $L$ is odd. For 6 or 7 loci, Eq. \ref{GlauberFormula} shows that we need expectations of 2-, 4- and 6-spin products. We have determined the 2- and 4-spin products above, and the mapping onto trees for computing the 6-spin product follows exactly the same logic as for the 4-spin product~\cite{SM-SixLoci}. More generally, when going from $L$ to $L+2$ loci, the only new unknown is the expectation of the product of all spins. Interestingly, the SD equations follow simple patterns \cite{SM-Rules}. Based on these patterns, we have written a computer program that takes as input the list of genetic positions of $L$ loci and computes the probability of all $L$-locus RIL genotypes~\cite{SM-Ccode}. Lastly, the approach is easily extended to the case where male and female recombination rates differ~\cite{SM-sexspecific}.

These exact rather than approximate probabilities of multi-locus genotypes could be used in a number of situations in which RIL probabilities are needed. For instance when building genetic maps, the ordering of markers relies on comparing likelihoods of multi-locus genotypes, generally approximated by products of pair-wise recombination rates over putatively adjacent loci~\cite{Lander_Green_1987}. The same approximation is routinely applied in algorithms for detection of quantitative trait loci using interval or composite interval mapping~\cite{CompositeMapping}. Similarly, when genotypes or haplotypes must be inferred or imputed because of missing information or because markers are not sufficiently dense~\cite{Servin_Stephens_2007}, determining the most likely assignment requires comparing multi-locus genotype probabilities. Moving beyond RILs, it is possible that our framework will unveil ways to perform calculations of multi-locus probabilities in more general population genetics contexts~\cite{Zanini_Neher_2012} where the main difficulty comes from having a potentially infinite number of generations. That situation arises when one is interested in fixation probabilities, steady-state multi-locus frequencies, or distribution times of the most recent common ancestor~\cite{Kingman_1982,Derrida_1999,Lohse_2011}

In 1931 Haldane and Waddington~\cite{Haldane_Waddington_1931} provided the exact 2-locus probabilities for successive generations (F2, F3, \ldots) based on recursion formulas from which they were able to extrapolate to RILs, {\it i.e.,} to an infinite number of generations. In the present work, we have instead directly treated the RIL situation, exploiting Eq. \ref{GlauberFormula} due to Glauber~\cite{Glauber_1963} and self-consistent equations of the Schwinger-Dyson type~\cite{Dyson_1949,Schwinger_1951}. {\it A posteriori}, it is quite surprising that these mathematical tools had not been used before to generalize the Haldane-Waddington formula. Perhaps just as surprising is their remarkable efficiency for solving this long-outstanding problem.

\begin{acknowledgments}
We thank D. de Vienne, C. Dillmann, M. Gromov, F. Hospital, S. Majumdar, N. Morozova and B. Servin for comments and references. AS and OCM acknowledge financial support from CNRS GDRE 513, ICTP and INRA.
\end{acknowledgments}



\beginsupplement
\newpage
\section*{\large Supplemental Material}

\section{Production of a Recombinant Inbred Line via Single Seed Descent}

Assume given two homozygous diploid parents $P_a$ and $P_A$ at the F0 generation. Without loss of generality, label the $L$ loci or markers of interest by 1, 2, \ldots $L$. We denote by $a_1$, $a_2$, \ldots $a_L$ the alleles of $P_a$, and by $A_1$, $A_2$, \ldots $A_L$ the alleles of $P_A$. The first generation of offspring or F1 individuals are produced by crossing $P_a$ and $P_A$ (Fig.~1 in Main Text and Fig. S1). Note that the F1 individuals are all identical and are heterozygous at each locus. Thus, the genotype of F1 individuals is $\{ a_1/A_1, a_2/A_2, \ldots a_L/A_L \}$. The construction of the next generation (F2) depends on whether individuals can be selfed or not. Most plants are hermaphrodites, the same individual being capable of producing both male and female gametes. Such plants can be selfed to produce offspring for the next generation, a process referred to as {\it single seed descent} (SSD) and illustrated in Fig.~1 in Main Text. For animals, it is necessary to cross brothers and sisters to produce offspring, and this is referred to as {\it sib} mating. The present work concerns SSD, the sib case being significantly more complex.

Each individual arising during the successive generations (F1, F2, F3, \ldots) has a genomic content corresponding to the union of two gametes produced within its progenitor: one via female meiosis and the other via male meiosis. These gametes often involve crossovers that mix alleles within chromosomes. For example, the F2 genotype in Fig.~1 in Main Text is $\{ a_1/a_1, a_2/A_2, A_3/a_3 \}$ and so the bottom chromosome is recombined for both intervals $(1,2)$ and $(2,3)$ due to the occurrence of a crossover in each interval. Recombination occurs during a meiosis if there are an odd number of crossovers between the 2 loci under consideration and as a result the interval $(1,3)$ of the example given is not recombinant. The probability that a recombination occurs between locus $i$ and locus $j$ is referred to as the (meiotic) recombination rate $r_{i,j}$ for that pair of loci. Crossovers form stochastically and their statistics has to be modeled. For pedagogical reasons, we follow standard practice and consider that female and male meioses are described by the same stochastic process and so that in particular female and male recombination rates are identical. Nevertheless, our framework is easily extended to the case of distinct female and male recombination rates (see Sections \ref{SM_TwoLoci} and \ref{SM_female_male} in Supplemental Material). Many models have been proposed to describe the statistics of crossover formation. In the simplest model, crossovers arise as independent events in each meiosis, a hypothesis due to Haldane \cite{Haldane_1919}. Other models take the crossovers to exhibit {\it interference} with close-by crossovers being very rare. Our framework allows any kind of crossover formation model to be treated since model dependencies are restricted to the probabilities $P(g)$ and $P(g')$ in Eq. 3 in Main Text. However, it is only in the case of no interference that the analytical calculations (using the mappings to trees) can be pushed very far.

If the $L$ loci are not physically linked, the calculation of the probabilities of genotypes at successive generations becomes trivial because the allelic content at each locus is passed on independently, corresponding to $r_{i,j}=1/2$. The whole complexity of finding the probabilities of multi-locus genotypes stems from the linkage between loci, \emph{i.e.}, $$r_{i,j} \ne 1/2$$ Thus, without loss of generality, we assume that all $L$ loci are on the same chromosome. After an F2 individual is produced, it is used to produce an F3 individual, which itself is used to produce an F4 individual, and so forth. If a locus becomes homozygous at one generation, it will remain {\it fixed} (neglecting mutations) in all future generations. If a locus is heterozygous at one generation, the probability that it will remain heterozygous at the next generation is 1/2. Thus, with the increase in the number of generations, more loci become homozygous and fixed. After a large number of generations, all alleles will become fixed (Fig.~1 in Main Text). If this SSD process is performed in parallel for a number of {\it lines} as illustrated in Fig.~S1, one obtains a population of recombinant inbred lines (RILs) where each genome is a homozygous mosaic of the two parental genomes. The different RIL individuals are inbred and any pair of loci may have recombined the alleles of the initial parents $P_a$ and $P_A$, thus the term RILs.

\begin{figure*}
\begin{center}
\includegraphics[width=14cm]{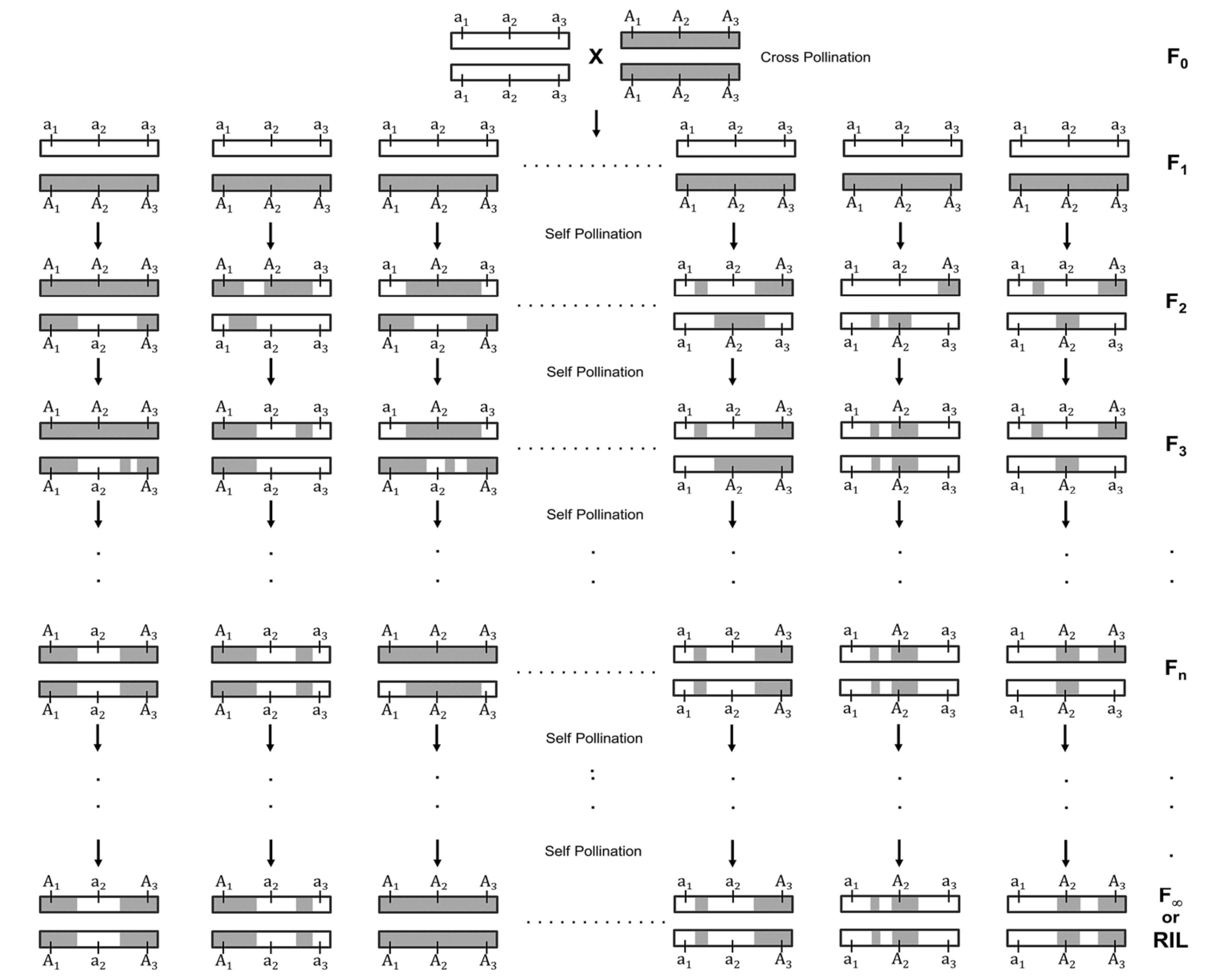}
\caption{{\bf Many Recombinant Inbred Lines produced in parallel using Single Seed Descent.} Two homozygous parents are crossed to produce the F1 generation of genetically identical individuals. Thereafter, at each generation, each plant produces one gamete via female meiosis and one gamete via male meiosis, and then these two gametes are fused to produce the genomic content of the individual of the next generation. Crossovers may arise during the meioses, leading to intra-chromosomal shuffling of allelic content. After enough generations, all loci become homozygous.}
\end{center}
\end{figure*}

\section{Rederiving the Haldane-Waddington formula via the new framework and the case of sex-specific recombination rates}
\label{SM_TwoLoci}

In 1918, Robbins \cite{Robbins_1918} determined the probabilities of RIL genotypes produced using single seed descent (SSD) for the case of 2 loci. In 1931, Haldane and Waddington \cite{Haldane_Waddington_1931} reconsidered the question using a simpler method and went on to solve the problem when using sib mating. Furthermore, Haldane and Waddington \cite{Haldane_Waddington_1931} also showed that the 2-locus RIL probabilities completely determine the 3-locus RIL probabilities. Here we show how our framework can simplify the derivation of the Haldane-Waddington formula for 2 loci in the SSD case, thus illustrating, albeit on a very simple case, the logic of the Schwinger-Dyson (SD) approach, approach that generalizes to many more loci.

To rederive the (2-locus) Haldane-Waddington formula, we start with the $2^2=4$ possible RIL genotypes, $\{ a_1/a_1, a_2/a_2 \}$, $\{ a_1/a_1, A_2/A_2 \}$, $\{ A_1/A_1, a_2/a_2 \}$, and $\{ A_1/A_1, A_2/A_2 \}$ where the top (bottom) allele specified at a locus is that on the chromosome generated during female (male) meiosis. In our spin notation, these homozygous genotypes are denoted as $\vec{S} = \{ 1,1 \}$, $\{ 1,-1 \}$, $\{ -1,1 \}$ and $\{ -1,-1 \}$, respectively. The RIL recombination rate $R$ is defined as the probability of having {\it recombinant} genotypes: $R = P(\{ 1,-1 \}) + P(\{ -1,1 \})$. $R$ is related to the expectation of the 2-spin product over all RIL genotypes with their respective probabilities via:
\begin{equation}
E [ S_1 S_2 ] = P(\{ 1,1 \}) + P(\{ -1,-1 \}) - P(\{ 1,-1 \}) - P(\{ -1,1 \}) = 1 - 2 R
\label{TwoSpinProduct}
\end{equation}

The difficulty in determining $R$ comes from the fact that producing RILs involves in principle an infinite number of generations. The heart of our method consists in transforming such an infinite process into a finite one based on self-consistent equations as follows. The probability of a RIL genotype is associated with sums over all possible meioses across generations F1, F2, \ldots \ leading to that RIL genotype. Now think of classifying these {\it trajectories} according to the genotype produced at generation F2. In our framework, we must calculate the probability of each F2 genotype and the contribution of associated trajectories to $E [ S_1 S_2 ]$. There are $4^2$ F2 genotypes and the probability of each is easy to compute.

Consider for instance the F2 genotype $\{a_1/A_1, A_2/A_2\}$ which occurs with probability $r(1-r)/4$ where $r=r_{1,2}$ is the recombination rate (for one meiosis) between the 2 loci. How much do trajectories passing through that F2 genotype contribute to $E [ S_1 S_2 ]$? Clearly, since the second locus is fixed to type ``A'', we have $S_2=-1$ necessarily. Furthermore, the first locus will fix to either $S_1=1$ or $S_1=-1$ with probability $1/2$ for each and summing these outcomes gives 0 for the expectation value of the 2-spin product. Thus trajectories passing through that F2 genotype contribute nothing to $E [ S_1 S_2 ]$. The same result will hold for all F2 genotypes that are heterozygous at one locus and homozygous at the other.

Consider then the F2 genotype $G=\{a_1/A_1, a_2/A_2\}$ arising with probability $P_G = (1-r)^2/4$. This genotype is identical to the F1 genotype, so its contribution is $P_G E [ S_1 S_2 ]$. The same result holds for the genotype $\{A_1/a_1, A_2/a_2\}$ because of the global invariance under exchange of all ``a''s for ``A''s and vice versa.

A bit more subtle is the case of the genotype $G'=\{a_1/A_1, A_2/a_2\}$ arising with probability $P_{G'} = r^2/4$. This case is {\it similar} to that of the F1 genotype except that the alleles at the second locus have been inverted. All trajectories produced from this genotype $G'$ can be mapped to those produced from the F1 genotype if we perform the {\it substitution} of the alleles at the second locus, exchanging ``a''s and ``A''s. The probabilities of these substituted trajectories will be the same as before the substitution but when we consider the contribution of genotype $G'$ in the RILs we have to also substitute $S_2 \rightarrow -S_2$. Thus the trajectories passing through the genotype $G'$ contribute the amount $-P_{G'} E [ S_1 S_2 ]$. The same strategy applies to the genotype $\{A_1/a_1, a_2/A_2\}$ for which the required substitution is $S_1 \rightarrow -S_1$.

Lastly, there are F2 genotypes that are homozygous at both loci. Their contribution to $E [ S_1 S_2 ]$ is easily read off since each locus is fixed, and in fact the RIL fixation has been accomplished in just one generation.

Adding up the contributions associated with all $4^2$ F2 genotypes gives the self-consistent equation:
\begin{equation}
E [ S_1 S_2 ] = \left[ \frac{(1-r)^2}{2} - \frac{r^2}{2} \right] \times E [ S_1 S_2 ] + \left[ 2(1-r)r \right] \times 0 + \left[ \frac{(1-r)^2}{2} - \frac{r^2}{2} \right]  \times 1
\label{SDTwoLoci}
\end{equation}
where we have ordered the terms according to F2 genotypes having 0, 1 and 2 fixed loci. This SD equation leads to $E [ S_1 S_2 ] (1 + 2 r) = 1 - 2r$ from which one obtains:
\begin{equation}
R = \frac{2r}{1+2r}
\end{equation}
{\it i.e.,} the Haldane-Waddington formula.

Robbins~\cite{Robbins_1918} and then Haldane and Waddington \cite{Haldane_Waddington_1931} also determined the probabilities of 2-locus SSD RIL genotypes in the case of sex-specific recombination rates, \emph{i.e.}, where the female and male recombination rates are different. Interestingly, that generalization does not affect much our framework, the only modification arises in the probabilities of the F2 genotypes. Denoting the female and male recombination rates between the 2 loci by $r^f$ and $r^m$, respectively, the generalization of Eq.~\ref{SDTwoLoci} along with obvious simplifications gives:
\begin{equation}
E [ S_1 S_2 ] = \left[ \frac{1-r^f-r^m}{2} \right] \times E [ S_1 S_2 ] + \left[ (1-r^f)r^m + r^f(1-r^m) \right] \times 0 + \left[ \frac{1-r^f-r^m}{2} \right]  \times 1
\end{equation}

It is interesting to note that although the individual $P(G)$s depend on both $r^f$ and $r^m$, the above SD equation depends only on the mean of $r^f$ and $r^m$ because the middle factor is multiplied by 0. This property is not general, and in particular, we will show later that it does not hold for 4 loci (see Section \ref{SM_female_male} in Supplemental Material). Solving for $R$ leads to:
\begin{equation}
R = \frac{r^f+r^m}{1+r^f+r^m}
\end{equation}
%

\section{Probabilities for 2 and 3 loci do not determine those for 4 loci}

In their 1931 paper, Haldane and Waddington \cite{Haldane_Waddington_1931} derived the formula for 2-locus RIL probabilities using recursions from one generation to the next and then took the limit of an infinite number of generations. Furthermore, they provided a simple mathematical trick involving standard algebra to obtain 3-locus RIL probabilities from 2-locus RIL probabilities (Fig.~2a in Main Text). Thus, those authors showed that the 2-locus RIL probabilities completely determine the 3-locus RIL probabilities. However, the simple trick of Haldane and Waddington does not extend to the case of 4 loci. We now elucidate this difference between going from 2 to 3 loci versus from 3 to 4 loci, and show that 2- and 3-locus RIL probabilities do {\it not} determine the 4-locus probabilities (Fig.~2 in Main Text).

Let us first calculate the probabilities of all ($2^3=8$) 3-locus RIL genotypes. One can use symmetries such as $P(a_1/a_1, a_2/a_2, A_3/A_3) = P(A_1,A_1, A_2/A_2 a_3/a_3)$ to formulate the problem in terms of 4 unknowns $Q(0,0)$, $Q(0,1)$, $Q(1,0)$ and $Q(1,1)$ defined in Fig.~2a in Main Text. For these 4 quantities, the binary entry 0 (respectively 1) denotes absence (respectively presence) of a recombination event in the corresponding interval (1 or 2). To determine these 4 unknowns, one requires 4 independent equations. A first independent equation is that the sum of the 4 probabilities equals 1. Furthermore, three additional equations are obtained from the 2-locus RIL probabilities using $R_{1,2}$, $R_{1,3}$ and $R_{2,3}$. Having as many independent equations as unknowns (the rank of the matrix constraining the 4 unknowns is 4), one concludes that the 2-locus probabilities uniquely determine the 3-locus probabilities. The mathematics behind the 3-locus case is provided in Fig.~2a in Main Text.

We next ask if all 2- and 3-locus probabilities similarly determine the 4-locus probabilities. There are $2^4=16$ 4-locus RIL genotypes. Again one can use symmetries to formulate the problem in terms of 8 unknowns (see $Q(0,0,0)$ etc. defined in Fig.~2b in Main Text). To determine these 8 unknowns, one needs 8 independent equations. As before, one equation follows from the fact that the sum of the 8 probabilities equals 1. Furthermore, there are 6 equations associated with 2-locus constraints ($R_{1,2}$, $R_{1,3}$, $R_{1,4}$, $R_{2,3}$, $R_{2,4}$, $R_{3,4}$). We need one more independent equation to solve for all 8 unknowns. It is tempting to use one of the equations based on 3-locus constraints (Fig.~2b in Main Text). However all those equations are consequences of the 2-locus constraints: they are {\it automatically} satisfied and provide no further constraints on the unknowns. In the 4-locus case the rank of the matrix constraining the 8 unknowns is at most 7 (Fig.~2b in Main Text) regardless of which of the 3-locus constraints are added since these follow from the 2-locus constraints. In conclusion, to obtain all 4-locus RIL probabilities, {\it one additional piece of information is needed} that is not incorporated in 2- or 3-locus RIL probabilities. Finding and exploiting this missing information has prevented researchers from extending the Haldane-Waddington result for over 80 years.

\begin{figure*}
\begin{center}
\includegraphics[width=14cm]{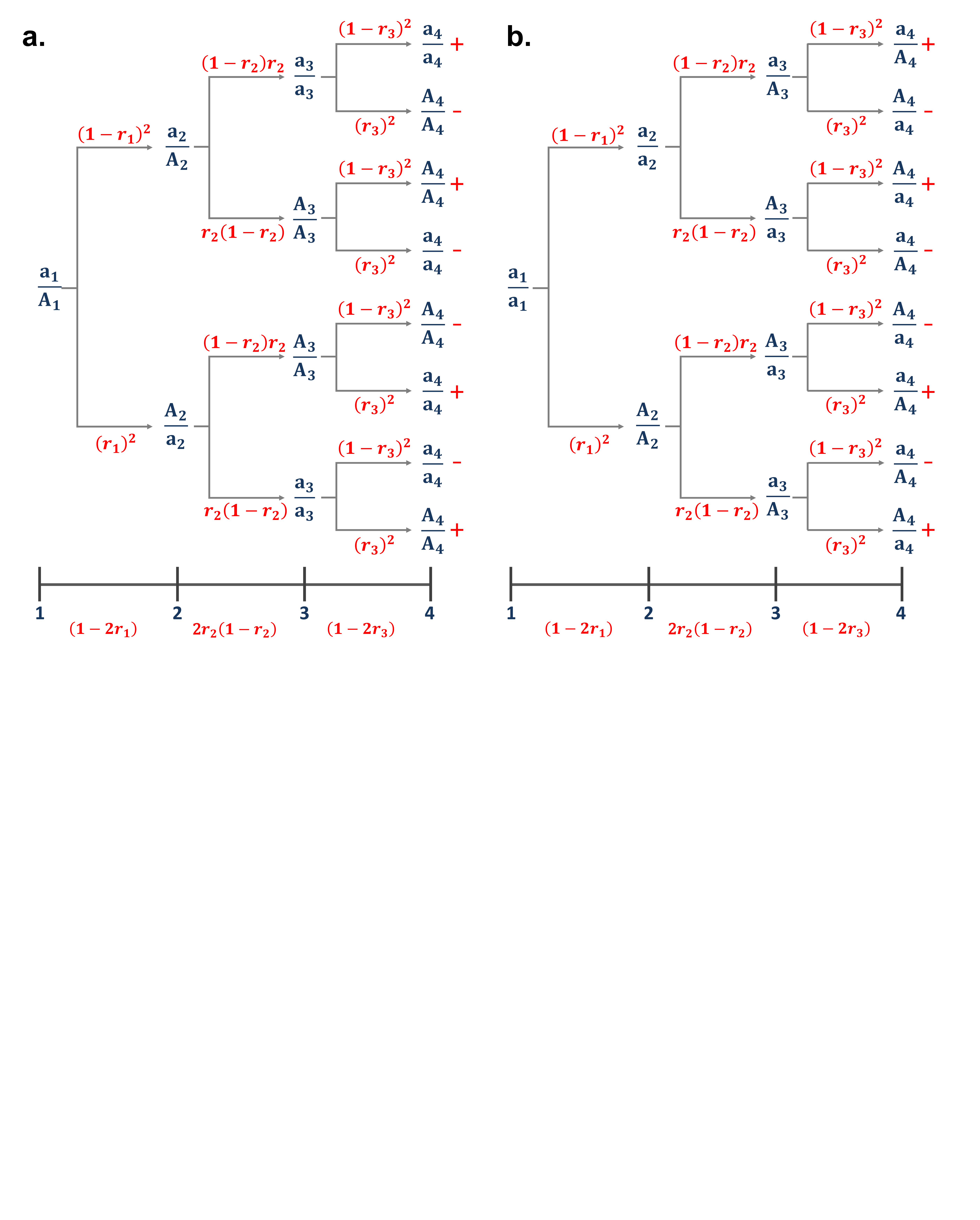}
\caption{{\bf Four-locus trees for classes of F2 genotypes with 2 fixed loci leading to non-vanishing contributions in the Schwinger-Dyson equation.} (a) $hhHH$ and (b) $HHhh$.}
\end{center}
\end{figure*}
\begin{figure*}
\begin{center}
\includegraphics[width=14cm]{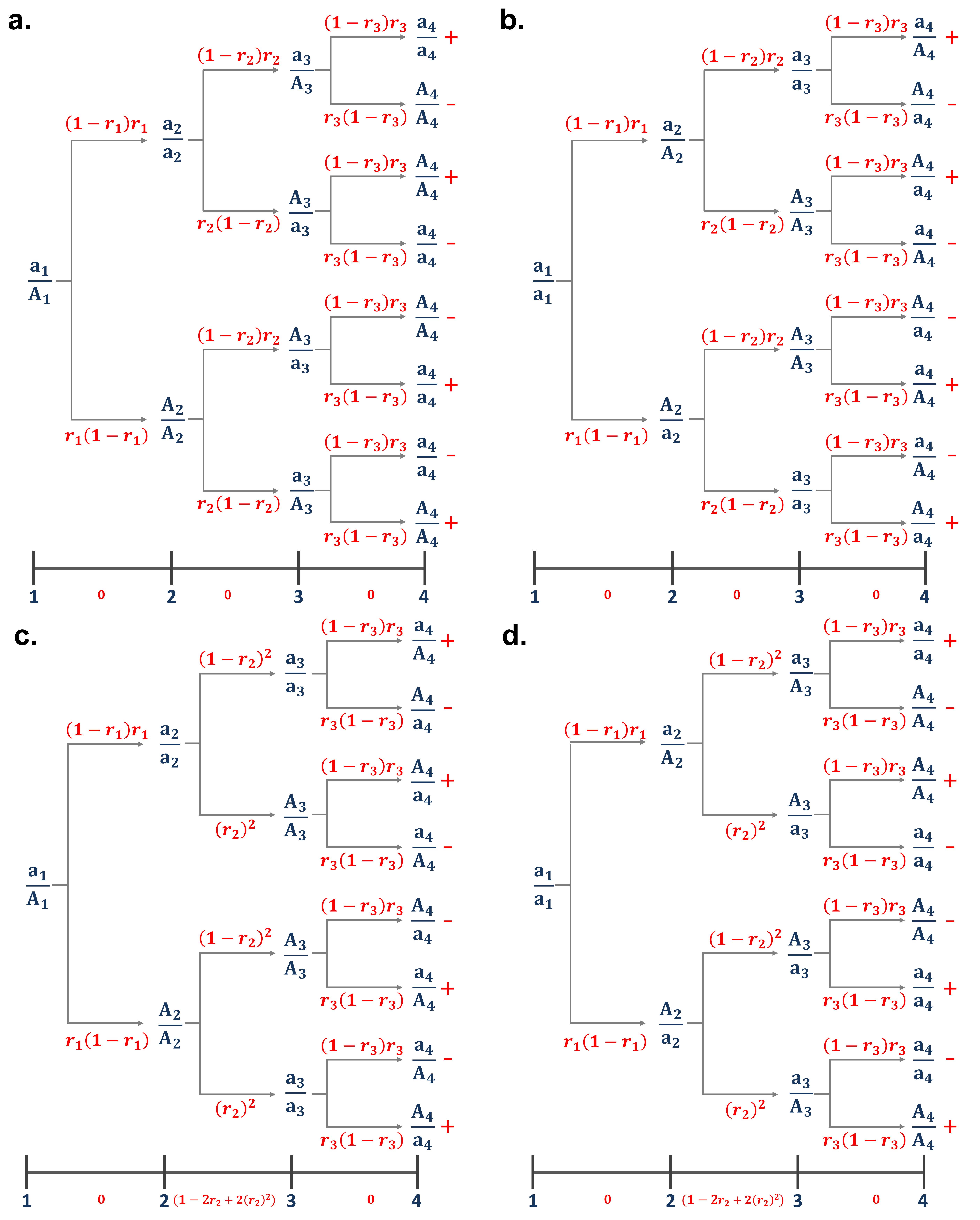}
\caption{{\bf Four-locus trees for classes of F2 genotypes with 2 fixed loci leading to vanishing contributions in the Schwinger-Dyson equation.} (a) $hHhH$, (b) $HhHh$, (c) $hHHh$ and (d) $HhhH$.}
\end{center}
\end{figure*}

\section{Mapping all F2 genotypes with 4 loci to trees}

To derive the SD Eq. 4 in Main Text for $E [ S_1 S_2 S_3 S_4 ]$, we classify the F2 genotypes according to whether they are homozygous (H) or heterozygous (h) at the different loci. There are $2^4$ such classes and each class contains $2^4$ genotypes because a homozygous (respectively, heterozygous) locus $i$ can have the allelic content $a_i/a_i$ or $A_i/A_i$ (respectively, $a_i/A_i$ or $A_i/a_i$). For illustration, consider all F2 genotypes belonging to the class $hhhh$. The first locus can be in the state $a_1/A_1$ or $A_1/a_1$, the second in the state $a_2/A_2$ or $A_2/a_2$, the third in the state $a_3/A_3$ or $A_3/a_3$, and so on. This succession of possibilities can be represented by a binary tree whose root is associated with the state of the first locus. Thus, there are 2 trees for the $hhhh$ class: one rooted on the $a_1/A_1$ state (Fig.~3 in Main Text) and the other on the $A_1/a_1$ state. These two trees are related to each other: one goes from one tree to the other via a global exchange of ``a''s into ``A''s and vice versa. In the Main Text, we mentioned this {\it exchange invariance} at the level of RIL probabilities but in fact it also holds for probabilities of genotypes at any generation of the RIL construction. Each F2 genotype $G$ can be identified with a path on its associated tree which goes from the root to a leaf of that tree. Furthermore, the probability of a genotype $G$, composed of its two gametes $g$ and $g'$, is the product of the following terms if crossovers arise without interference: a factor $1/4$ for the root node and a factor $(1-r)^2$, $(1-r)r$, $r(1-r)$ or $r^2$ for each interval between adjacent loci depending on whether the interval is recombinant or not for $g$ and for $g'$. Finally, each genotype $G$ comes with a sign, denoted here by $\text{sign}(G)$, which arises from the substitution rules (see Main Text).

It is easy to enumerate all the classes to cover: $hhhh$, $hhhH$, $hhHh$, $hhHH$, and so on. Each class gives rise to two trees. However, just as in the example considered above, these two trees are related by the exchange invariance under swapping of ``a''s and ``A''s. Thus, it is enough to consider one tree per class and to multiply its contribution by 2 in the SD equation. Thus, without loss of generality, we show in all our figures only those trees which have at locus 1 of their female chromosome the $a_1$ allele. With this first simplification, the number of trees to be considered is $2^4$.

A second major simplification arises by noting that each class is associated with a different multi-spin product to average. For instance, if one considers the terms on the right-hand side of the SD Eq. 4 in Main Text, $hhhh$ is associated with $E[S_1 S_2 S_3 S_4]$, $hhhH$ with $E[S_1 S_2 S_3]$, $hhHh$ with $E[S_1 S_2 S_4]$, and so on. Because the probabilities of RIL genotypes are invariant under $\vec{S} \rightarrow -\vec{S}$, expectations are also invariant. Then if there is an odd number of spins in a spin-product, its expectation value vanishes. As a consequence, amongst the $2^4$ classes, we only need to consider those with 0, 2 and 4 loci of type $H$.

Consider the tree for the class $hhhh$ rooted at $a_1/A_1$ (Fig.~3 in Main Text). The {\it sign} carried by a genotype is specified on the leaf of the path representing $G$ on its tree. By summing $\text{sign}(G) P(G)$ over all genotypes $G$ belonging to this tree, and multiplying by 2 to take into account the other tree for this class ({\it i.e.}, the tree rooted at $A_1/a_1$), we obtain the factor in the right-hand side of the SD equation associated with $E[S_1 S_2 S_3 S_4]$. To derive the formula for this sum, we start with the right-most of the three intervals and collect the paths into pairs that differ only in this last interval. This pools together contributions of double recombinants and double non-recombinants with opposite signs, leading to the factor $(1-r_{3})^2 - r_{3}^2 = 1 - 2 r_{3}$ and a sign that depends on the pair. The factor for the third interval is given at the bottom of the tree in Fig.~3 in Main Text. An important point is that this factor $1-2 r_3$ is common to all pairs of paths which differ only in the last interval, and so these pairs can be identified with shortened paths restricted to just the first two intervals. This property allows us to {\it iterate} the procedure. Thus we consider now all (shortened) paths covering just the first two intervals and pair these up if they differ only on the second interval. Again, the pairing requires pooling the contributions of double recombinants and double non-recombinants. Because the two paths to be added in the pair both have the same sign (which was not true for the third interval), the common factor for the second interval is $(1-r_2)^2 + r_2^2$ (Fig.~3 in Main Text). This pooling leaves us with just two shortened paths of one segment with opposite signs for the first interval (Fig.~3 in Main Text). Thus, adding the contributions of these two paths we obtain the factor for the first interval $(1-r_1)^2 - r_1^2 = 1 - 2 r_1$ (Fig.~3 in Main Text). One final factor must be included: the probability of having the first locus in the given ($a_1/A_1$) state, {\it i.e.,} $1/4$. The resulting product is this tree's contribution to $E[ S_1 S_2 S_3 S_4]$, coming from its 8 F2 genotypes. There are also the other 8 F2 genotypes of the class $hhhh$ associated with a tree rooted on $A_1/a_1$ which leads to exactly the same result as can be seen either by direct calculation or by using the previously mentioned exchange invariance under global swaps of ``a''s and ``A''s. Putting all this together, the factor in front of $E[ S_1 S_2 S_3 S_4 ]$ on the right-hand side of Eq. 4 in Main Text is:
\begin{equation}
A_{1,1,1,1} = \frac{(1 - 2 r_{1})  \left[ (1-r_{2})^2 + r_{2}^2 \right] (1 - 2 r_{3})}{2}
\end{equation}
where the indices of $A_{1,1,1,1}$ refer to the powers $n_i$ arising in the associated spin product $E [ S_1^{n_1} S_2^{n_2} S_3^{n_3} S_4^{n_4} ]$.

Suppose one repeats the calculation that led to $A_{1,1,1,1}$ but replaces the male chromosome in each $G$ by a modified one where all ``a''s have been exchanged for ``A''s and vice versa. This transformation takes one from a heterozygous $G$ to a homozygous $G'$. Interestingly, this transformation affects neither the probabilities arising in each interval nor the signs ($\text{sign}(G) = \text{sign}(G')$). Thus, $A_{0,0,0,0} = A_{1,1,1,1}$. Furthermore, it is easy to see that this invariance applies to any of the F2 genotypes. As a result, for any choices of the $n_i$, ($n_i = 0$ or 1), $A_{n_1,n_2,n_3,n_4} = A_{1-n_1,1-n_2,1-n_3,1-n_4}$, providing the third major simplification and reduction in the set of trees to be considered.

Finally we are left with the mixed cases where $G$ has two homozygous loci and two heterozygous loci. Let us begin with the class $hhHH$ which determines the factor $A_{1,1,0,0}$ (Fig.~S2a in Supplemental Material). The contributions from the associated F2 genotypes can be combined just as in the calculation of $A_{1,1,1,1}$: the third interval again leads to the factor $(1 - 2 r_{3})$; the second interval has one recombinant gamete and one non-recombinant gamete, leading to the factor $2 (1-r_{2}) r_{2}$; and finally the first interval again leads to the factor $(1 - 2 r_{1})$. Using the fact that two trees contribute to the class $hhHH$ and the invariance result from the previous paragraph for deducing $A_{0,0,1,1,}$ (Fig.~S2b in Supplemental Material), we obtain:
\begin{equation}
A_{1,1,0,0} = A_{0,0,1,1} = \frac{(1 - 2 r_{1}) ( 2 (1-r_{2}) r_{2} ) (1 - 2 r_{3})}{2}
\end{equation}

The other mixed cases lead to an even simpler result. Consider for instance the class $hHhH$ and the associated tree (Fig.~S3a in Supplemental Material). When paths differing only on the last interval are paired, the interval factor for each genotype is $(1-r_{3})r_{3}$ but the signs are opposite and so the sum vanishes. The same is true for the remaining classes ($HhHh$, $hHHh$, and $HhhH$), and thus $A_{1,0,1,0}=A_{0,1,0,1}=A_{1,0,0,1}=A_{1,0,0,1}=0$ (Fig.~S3 in Supplemental Material).

Thus, collecting the results from all classes of F2 genotypes leads to the 4-locus SD equation (Eq. 4 in Main Text) where the expectation of the 4-spin product arises on both sides of this equation. Using the formula for the expectation of the 2-spin product: $E[ S_i S_j ] = 1 - 2 R_{i,j} = (1 - 2 r_{i,j})/(1 + 2 r_{i,j})$ in Eq. 4 of Main Text one obtains the expectation of the 4-spin product as:
\begin{equation}
E[ S_1 S_2 S_3 S_4 ] = \frac{F + A}{1 - F}
\end{equation}
where the terms $F$ and $A$ in this equation are given by:
\begin{equation}
F = \frac{(1 - 2 r_1)  ( (1-r_2)^2 + r_2^2 ) (1 - 2 r_3)}{2} \textnormal{ and }
A = \frac{(1 - 2 r_1) (2 (1-r_2) r_2 ) (1-2 r_3) (1 - 4 r_1 r_3)}{(1 + 2 r_1)(1 + 2 r_3)}.
\end{equation}
The probabilities of RIL genotypes can be obtained by substituting the expectations of the 2-spin products given by $E[ S_i S_j ] = 1 - 2 R_{i,j} = (1 - 2 r_{i,j})/(1 + 2 r_{i,j})$ and the expectation of the 4-spin product from Eq. 12 into Eq. 2 in Main Text. For instance, the probability of the four-locus genotype $\{ a_1/a_1, a_2/a_2, a_3/a_3, a_4/a_4 \}$ is
\begin{equation}
P(\{ 1,1,1,1 \}) = \frac{1 + \sum_{i<j} \frac{1 - 2 r_{i,j}}{1 + 2 r_{i,j}}  + \frac{F + A}{1-F}}{16}
\end{equation}
%

\section{Useful properties for simplifying the derivation of the Schwinger-Dyson equations}
\label{SM_rules}

In the 4-locus case we made use of a number of identities to reduce the number of F2 genotypes that had to be considered in the SD equation. Here we make such properties explicit for the general case of any number of loci and also introduce one additional invariance.

\noindent {\bf Rule~1:} For each class of F2 genotypes (denoted by a succession of $L$ letters in \{$H$,$h$\}), there are two associated trees: one with allele $a_1$  and the other with allele $A_1$ at locus 1 for the female chromosome. In fact the two trees lead to the same contribution to the SD equation. So, in practice one can force the allele at locus 1 for the female chromosome to be $a_1$, reducing by a factor 2 the number of trees to be considered.

\noindent {\bf Rule~2:} For a given class of F2 genotypes, the spin product $E [ S_1^{n_1} S_2^{n_2} \ldots S_L^{n_L} ]$ generated in the SD equation has $n_i=1$ if the locus $i$ is of type $h$ and $n_i=0$ if the locus $i$ is of type $H$. The number of spins in the spin product is then equal to the number of heterozygous loci in the class. Given the invariance of expectations values under the change of sign of all spins, the expectation value of a $k$-spin product vanishes when $k$ is odd. Thus a second simplification consists in keeping only the classes of genotypes having an even number of $h$'s, again reducing by a factor of 2 the number of trees to be considered.

\noindent {\bf Rule~3:} A further useful property is {\it chromosome choice invariance}. Consider exchanging ``a''s and ``A''s on just {\it one} of the chromosomes of an F2 genotype. In terms of meiosis, this corresponds to exchanging the two (F1) parental chromosomes when producing that gamete. In terms of the classes of F2 genotypes, it leads to the global swap of $H$s and $h$s, taking one class to a transformed one. A tree of the first class is transformed to a tree of the second class but the probabilities and signs are left invariant. However at the level of spin products, the transformation changes $n_i=1$ into $n_i=0$ and vice versa. As a result, factors in the SD equation come in equal pairs, for example $A_{1,0,0,1,0,0} = A_{0,1,1,0,1,1}$, reducing again by a factor 2 the number of trees to be considered. Note that if one applies chromosome choice invariance successively to both the male and the female chromosomes, {\it all} ``a''s and ``A''s are exchanged; then the class considered (list of $H$s and $h$s) is invariant but the allele at the first locus changes from $a_1$ to $A_1$, leading to Rule~1 which is thus a special case of Rule~3. Collecting these results, there are always four trees that produce exactly the same factors (albeit multiplying different expectation values in the SD equation), these trees being rooted on $a_1/A_1$, $A_1/a_1$, $a_1/a_1$ and $A_1/A_1$.

\noindent {\bf Rule~4:} For a class of F2 genotypes to lead to a non-zero contribution in a SD equation, both the $h$ loci and the $H$ loci must come in adjacent pairs. To see why this is the case, consider a class of F2 genotypes in which there is a block of adjacent $H$ loci, delimited by $h$ loci, and let $G$ be one genotype in this class. In the left interval bounding this block, one of the chromosomes of $G$ is recombinant, the other not. The same property holds for the right interval bounding this block. Consider now the F2 genotype $G'$ identical to $G$ in terms of crossover locations except that for these two intervals we exchange which is the chromosome (female or male) that is recombinant. This transformation does not affect the  probability of the genotype, but $\text{sign}(G') = - \text{sign}(G)$ if and only if the size of the $H$ block is odd. The contribution of $G'$ thus cancels exactly that of $G$ in such a situation. This still holds if the block of $H$s has only one interval bounding it ({\it i.e.}, it goes to an end of the chromosome). And by symmetry, the whole argument can be repeated when considering blocks of $h$s instead of blocks of $H$s. This fourth rule was used while studying the 4-locus case, but it is completely general and greatly reduces the number of classes to consider when there are many loci.

\begin{figure*}
\begin{center}
\includegraphics[height=24cm]{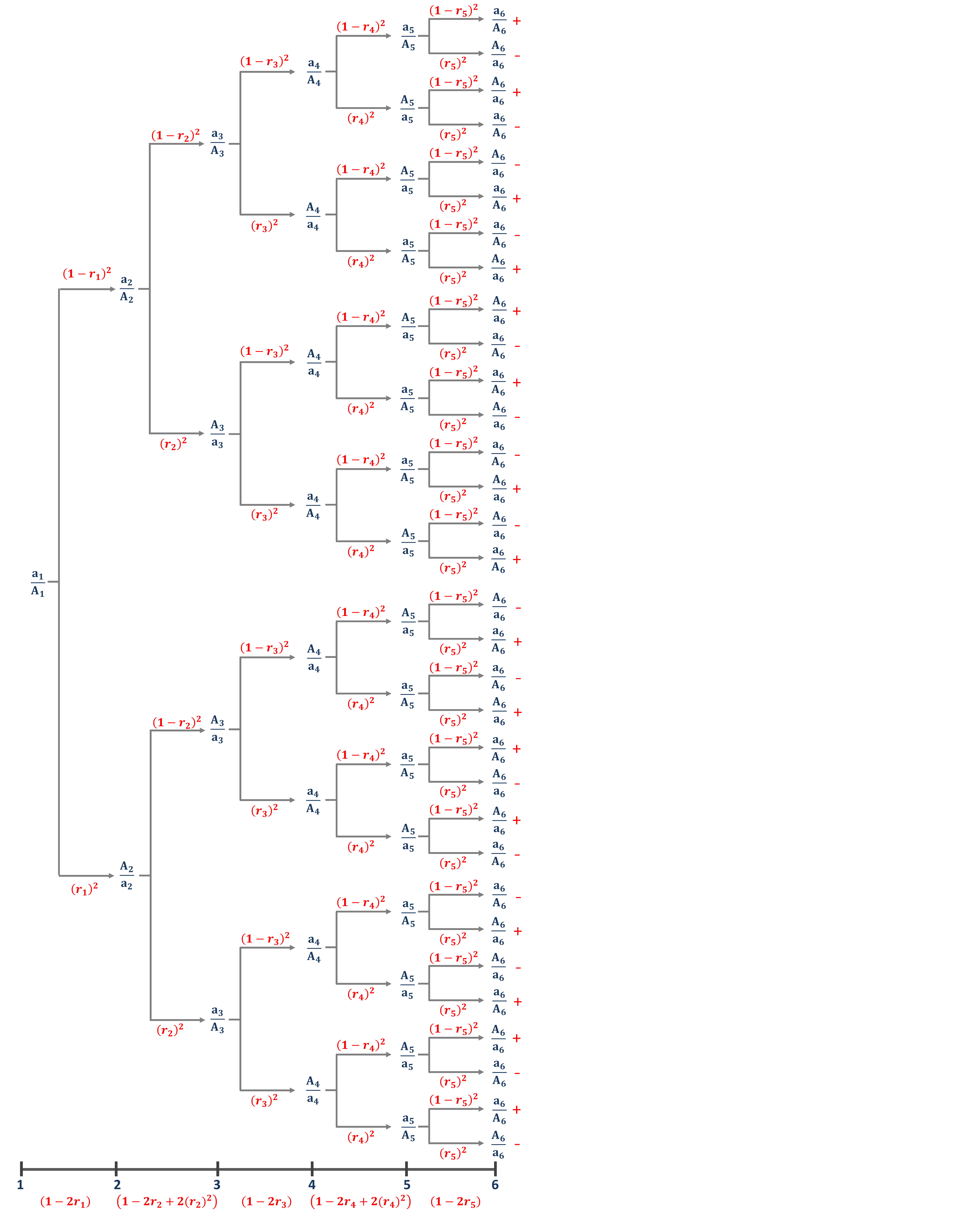}
\caption{{\bf Six-locus tree for the class $hhhhhh$ of F2 genotypes.}}
\end{center}
\end{figure*}
\begin{figure*}
\begin{center}
\includegraphics[height=24cm]{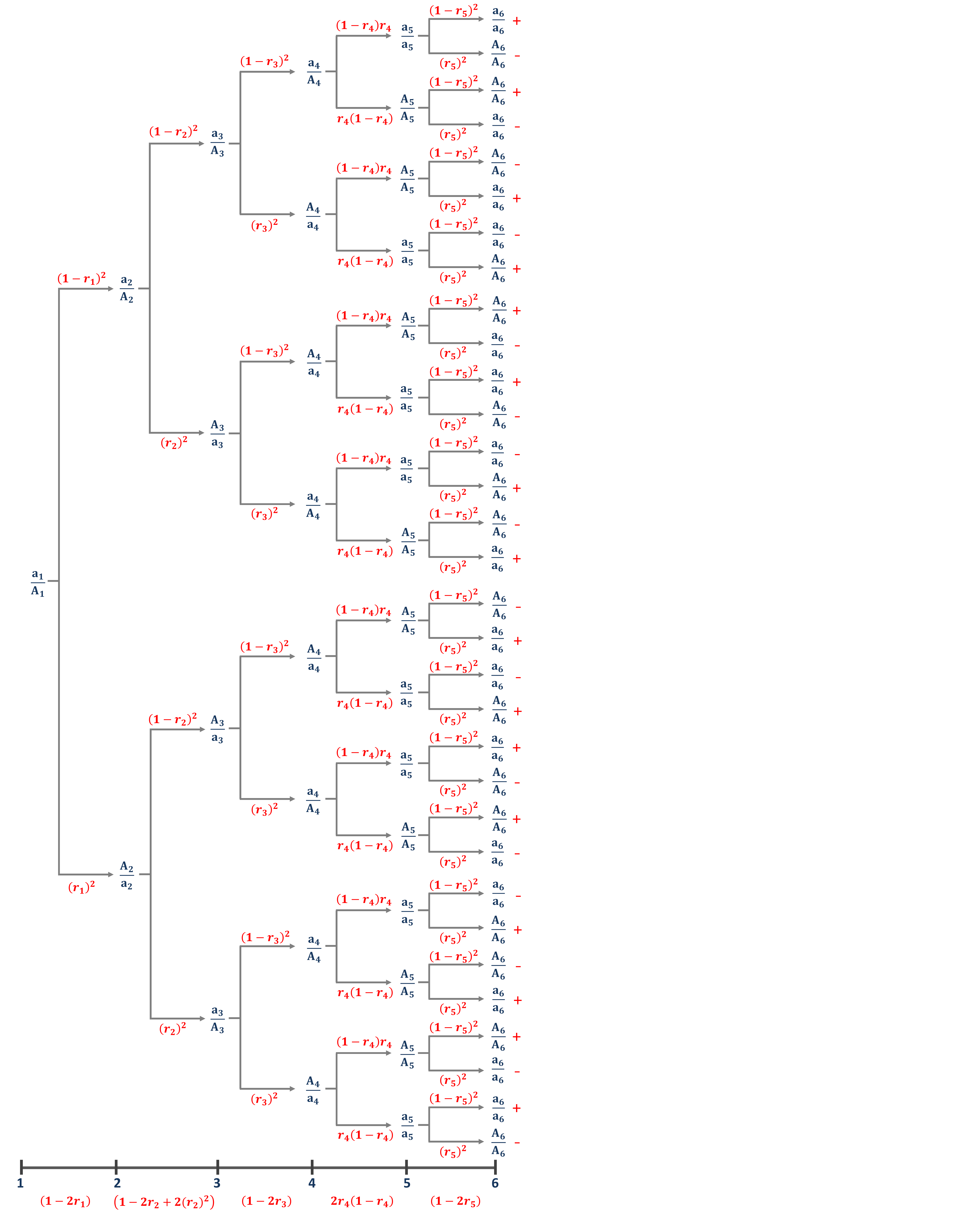}
\caption{{\bf Six-locus tree for the class $hhhhHH$ of F2 genotypes.}}
\end{center}
\end{figure*}
\begin{figure*}
\begin{center}
\includegraphics[height=24cm]{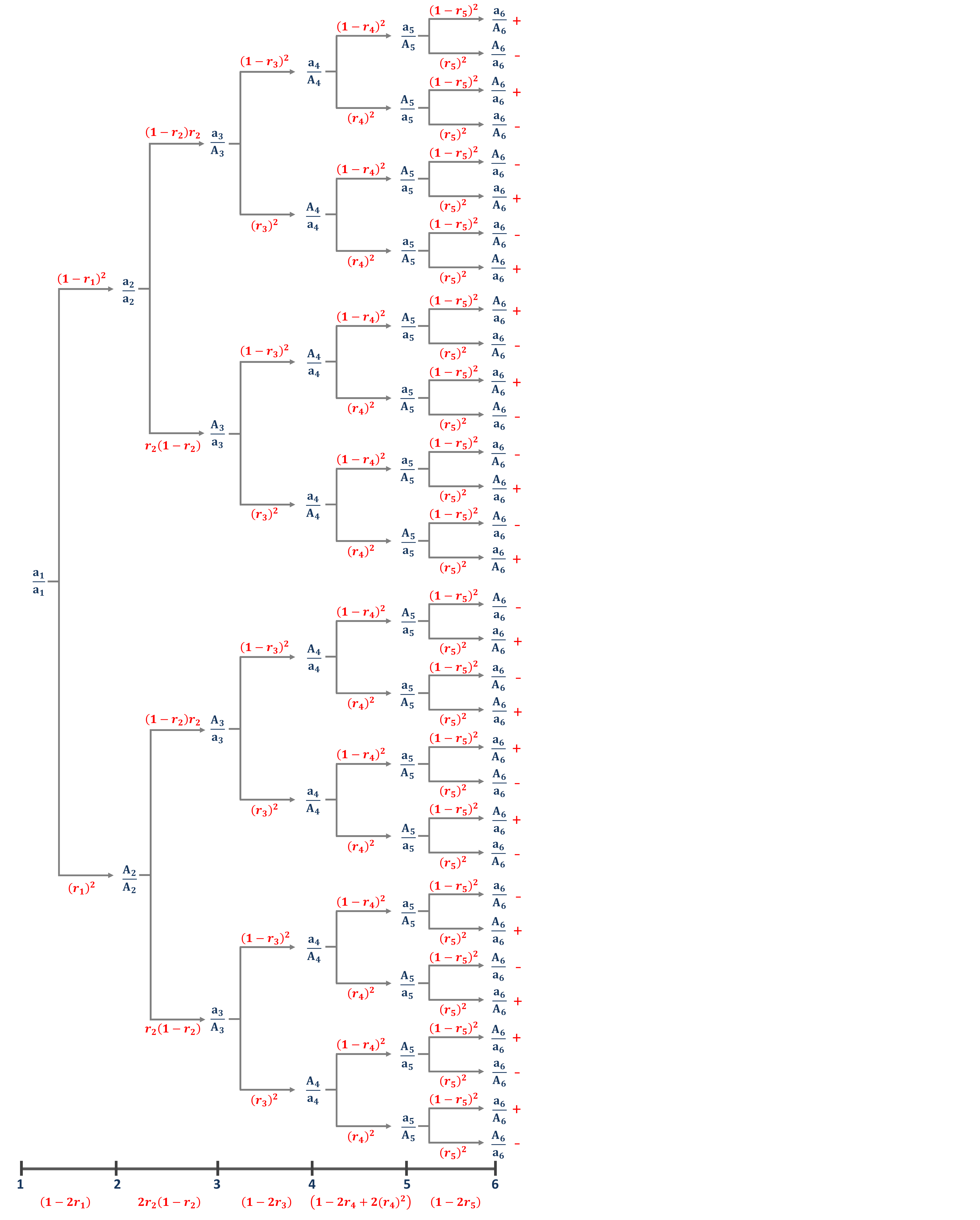}
\caption{{\bf Six-locus tree for the class $HHhhhh$ of F2 genotypes.}}
\end{center}
\end{figure*}
\begin{figure*}
\begin{center}
\includegraphics[height=24cm]{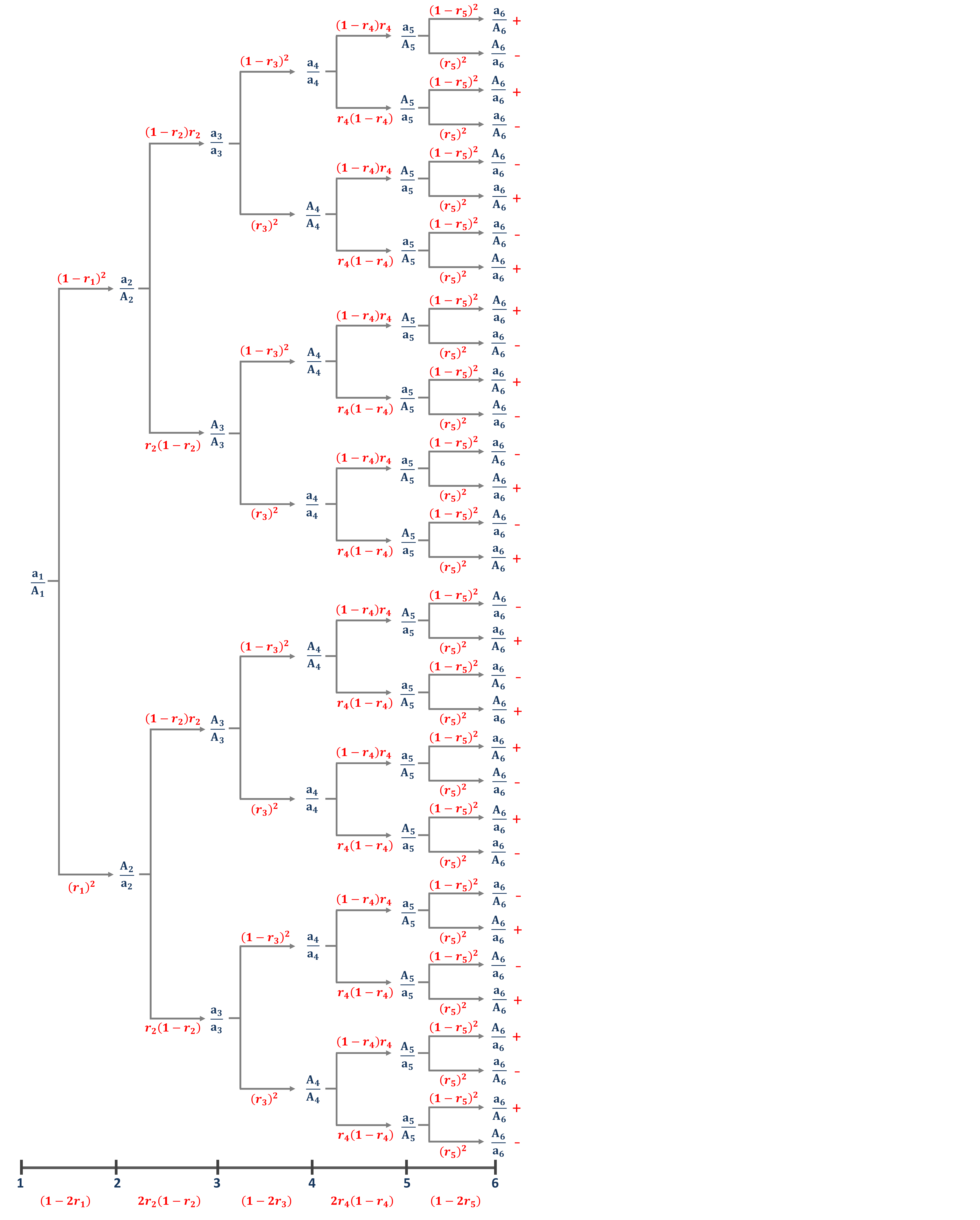}
\caption{{\bf Six-locus tree for the class $hhHHhh$ of F2 genotypes.}}
\end{center}
\end{figure*}

\section{The Schwinger-Dyson equations for 6 loci and beyond}

Using the SD framework along with the simplification rules listed in Section \ref{SM_rules} in Supplemental Material, the following four classes must be considered in the case of 6 loci: $hhhhhh$ that gives the factor for the $E[ S_1 S_2 S_3 S_4 S_5 S_6 ]$ and $1$ terms; $hhhhHH$ that gives the factor for the $E[ S_1 S_2 S_3 S_4 ]$ and $E[ S_5 S_6 ]$ terms; $HHhhhh$ that gives the factor for the $E[ S_3 S_4 S_5 S_6 ]$ and $E[ S_1 S_2 ]$ terms; $hhHHhh$ that gives the factor for the $E[ S_1 S_2 S_5 S_6 ]$ and $E[ S_3 S_4 ]$ terms.

Consider the class $hhhhhh$ with the tree rooted at $a_1/A_1$ (Fig.~S4 in Supplemental Material). Just as in the 4-locus case, an F2 genotype associated with this tree corresponds to a path from the root of the tree to one of the leaves of the tree. To determine the sum of $\text{sign}(G) P(G)$ over the F2 genotypes, we again start at the right-most (fifth) interval and collect into pairs the paths that differ only for that last interval. The calculation is identical to that performed for the 4-locus case and one obtains the factor $(1-r_5)^2 - r_5^2$. Similarly (and not surprisingly in view of how the calculation proceeded in the 4-locus case), the fourth interval leads to the factor $(1-r_4)^2 + r_4^2$. After having treated those two intervals, we see that the remaining paths correspond to a 4-locus tree that is identical with the one for the $hhhh$ class on loci 1 to 4 (Fig.~3 in Main Text). Thus the 6-locus tree for the $hhhhhh$ class gives a factor that is the product of $(1-2r_5)$, of $\left[ (1-r_4)^2 + r_4^2 \right]$, and of the previously derived factor for the tree for the $hhhh$ class on loci 1 to 4, so that
\begin{equation}
A_{1,1,1,1,1,1} = \frac{(1 - 2 r_1)  \left[ (1-r_2)^2 + r_2^2 \right] (1 - 2 r_3) \left[ (1-r_4)^2 + r_4^2 \right] (1 - 2 r_5)}{2}
\end{equation}

Consider next the class $hhhhHH$ and its tree rooted at $a_1/A_1$ (Fig.~S5 in Supplemental Material). Proceeding as before, we pool together the paths that differ only in the last interval, leading to the common factor $(1-r_5)^2 - r_5^2$. Moving on to the fourth interval, we see that it takes one from a locus of type $h$ to a locus of type $H$, leading to the factor $2(1-r_4)r_4$. After this, the remaining tree is identical with the one for the $hhhh$ class on loci 1 to 4 (Fig.~3 in Main Text), just as in the previous paragraph. From this we conclude that the 6-locus tree for the $hhhhHH$ class gives a factor that is the product of $(1-2r_5)$, of $\left[ 2(1-r_4)r_4 \right]$, and of the previously derived factor for the tree for the $hhhh$ class on loci 1 to 4, and thus
\begin{equation}
A_{1,1,1,1,0,0} = \frac{(1 - 2 r_1)  \left[ (1-r_2)^2 + r_2^2 \right] (1 - 2 r_3) \left[ 2(1-r_4)r_4 \right] (1 - 2 r_5)}{2}
\end{equation}

Moving on to the class $HHhhhh$, the pooling over the last two intervals of the tree (Fig.~S6 in Supplemental Material) leads to the same factors as those obtained for the $hhhhhh$ class. After this, the remaining tree is identical with the one for the $HHhh$ class on loci 1 to 4 (Fig.~S2b in Supplemental Material). Of course, the same result could also have been obtained from the formula for the class $hhhhHH$ by taking the convention that loci are ordered from right to left rather than from left to right. This gives
\begin{equation}
A_{0,0,1,1,1,1} = \frac{(1 - 2 r_1)  \left[ 2(1-r_2)r_2  \right] (1 - 2 r_3) \left[ (1-r_4)^2 + r_4^2  \right] (1 - 2 r_5)}{2}
\end{equation}

Finally, consider the tree associated with the last class $hhHHhh$ (Fig.~S7 in Supplemental Material). The pooling over the last two intervals of the tree leads to the factors $(1-2r_5)$ and $2(1-r_4)r_4$. After this, the remaining tree is identical to the one for the $hhHH$ class on loci 1 to 4 (Fig.~S2a in Supplemental Material). As a result,
\begin{equation}
A_{1,1,0,0,1,1} = \frac{(1 - 2 r_1)  \left[  2(1-r_2)r_2  \right] (1 - 2 r_3) \left[ 2(1-r_4)r_4 \right] (1 - 2 r_5)}{2}
\end{equation}

Collecting all terms, one obtains the 6-locus SD equation:
\begin{equation}
\begin{split}
E [ S_1 S_2 S_3 S_4 S_5 S_6 ] &= A_{1,1,1,1,1,1} (E [ S_1 S_2 S_3 S_4 S_5 S_6 ] + 1) + A_{1,1,1,1,0,0} (E [ S_1 S_2 S_3 S_4 ] + E [ S_5 S_6 ] ) \\
                              &+ A_{0,0,1,1,1,1} (E [ S_3 S_4 S_5 S_6 ] + E [ S_1 S_2 ] ) + A_{1,1,0,0,1,1} (E [ S_1 S_2 S_5 S_6 ] + E [ S_3 S_4 ] )
\end{split}
\end{equation}
Note that these equations can be generalized to the case where recombination rates differ between male and female meiosis (see Section \ref{SM_female_male} and Fig.~S8 in Supplemental Material).

The patterns found in all these equations are easily extended to any number of loci. The SD equation for $E [ S_1 S_2 \ldots S_L ]$ can be written in terms of factors $A_{n_1,n_2, \ldots n_L}$ and associated expectations of multi-spin products where the indices of these factors must satisfy the constraint of Rule~4 in Section \ref{SM_rules} in Supplemental Material: 0s and 1s must come in adjacent pairs. For each such $A_{n_1,n_2, \ldots n_L}$, there is a global factor of $1/2$, a factor for each block of a given type (block of 0s or block of 1s), and one factor for each interval {\it connecting} blocks. The factor for connecting two blocks is $\left[  2(1-r)r  \right]$, $r$ being the recombination rate in that connecting interval. The factor {\it within} a block is a product over all of its intervals, alternating between $(1 - 2 r)$ terms and $\left[ (1-r)^2 + r^2  \right]$ terms and ending with a $(1 - 2 r)$ term because the number of intervals is odd. These results show that the SD equations can be written down {\it automatically} for any number of loci.

\section{Generalizing the formulas to sex-specific recombination rates}
\label{SM_female_male}

We saw how to generalize the standard Haldane-Waddington 2-locus formula for $R$ to situations where the female and male meiotic recombination rates $r^f$ and $r^m$ differ (see Section \ref{SM_TwoLoci} in Supplemental Material). Interestingly, it is also possible to generalize {\it all} our $L$-locus formulas to such a situation as follows.

First, the Glauber formula (Eq. 1 in Main Text) that gives the probabilities of the RIL genotypes in terms of expectation values of spin products is unchanged because it does not involve recombination rates and even less sex-specificity. Second, moving on to the SD equations, sex-dependence arises only at the level of the probabilities of gametes, {\it i.e.,} through the probabilities $P(g)$ and $P(g')$ (Eq. 3 in Main Text). The probabilities $P(g)$ and $P(g')$ must be modified but otherwise the logic is the same as in the sex-independent case. Specifically, one considers classes of F2 genotypes according to whether the successive loci are homozygous ($H$) or heterozygous ($h$). One maps these genotypes to binary trees as before to obtain a factor that multiplies an expectation value in the SD equation. That factor is a product of terms, one for each interval between adjacent loci. If, in the sex-independent case, an interval contributed the factor $(1-r)^2 - r^2$ (which simplifies to $1-2r$), it will now contribute $(1-r^f)(1-r^m)-r^f r^m$ (which simplifies to $(1 - r^f - r^m)$). If an interval contributed $(1-r)^2 + r^2$ in the sex-independent case, it will now contribute $(1-r^f)(1-r^m) + r^f r^m$. If an interval contributed $2(1-r)r$ in the sex-independent case, it will now contribute $(1-r^f)r^m + r^f(1-r^m)$. However, this is not the end of the story: in the sex-independent case, a large number of trees were discarded because one of the intervals led to the factor 0 (Fig.~S3 in Supplemental Material). For instance, for the class $hHhH$ in the sex-independent case, when one does the pooling of pairs in the right-most interval, one is led to $(1-r)r - r(1-r)$ which shows that the tree can be ignored (Fig.~S3a in Supplemental Material). However, in the sex-specific case, that factor becomes $(1-r^f)r^m - r^f(1-r^m) = r^m -r^f$ which has no reason to vanish. Going back to the rules listed in Section \ref{SM_rules} in Supplemental Material, it transpires that Rule~4 requires exchanging female and male meiosis. Thus, for sex-specific rates, this last rule and its associated simplifications have to be abandoned.

To illustrate the changes required for sex-specific recombination rates, consider the SD equation for $E [ S_1 S_2 S_3 S_4 ]$. The right-hand side of that SD equation contains one factor multiplying $E [ S_1 S_2 S_3 S_4 ]$ (the self term), factors for all of the $E [ S_i S_j ]$  terms (for {\it any} pair $(i,j)$ of distinct loci, not just the (1,2) and (3,4) pairs found in the sex-independent case), and finally a factor multiplying 1 (no associated expectation). The fact that no other expectation values contribute is due to Rule~2 in Section \ref{SM_rules} in Supplemental Material. Rule~3 in Section \ref{SM_rules} in Supplemental Material implies that the factor in front of $E [ S_1 S_2 S_3 S_4 ]$ is the same as in front of $1$, and also that the factor in front of $E [ S_i S_j ]$ is the same as in front of $E [ S_k S_l ]$ where $i$, $j$, $k$, and $l$ are all distinct. As an example, to obtain the factor multiplying $E [ S_2 S_4 ]$, we need to consider the F2 genotypes that are heterozygous at loci 2 and 4 and homozygous at loci 1 and 3. This class of genotypes gives two trees, one rooted at $a_1/a_1$ and other at $A_1/A_1$. By Rule~1 in Section \ref{SM_rules} in Supplemental Material, these contribute equally to the SD equation so it is sufficient to consider the first tree. For the sex-independent case, this tree (associated with class $HhHh$) satisfies all the rules listed in Section \ref{SM_rules} in Supplemental Material except the last rule, so it vanishes when $r^f = r^m$ (Fig.~S3b in Supplemental Material). The calculation of factors of this tree for the {\it sex-specific} case ($r^f \ne r^m$) are given in Fig.~S8 in Supplemental Material to which must be taken into account the factor $1/4$ for the root of this tree. Putting these factors together, we conclude that the term multiplying $E[S_2 S_4]$ in the SD equation is:
\begin{equation}
A_{0,1,0,1} = \frac{( r_1^m - r_1^f ) [ r_2^m + r_2^f - 2 r_2^m r_2^f ] ( r_3^m - r_3^f ) }{2}
\end{equation}
in the sex-specific case.

\begin{figure*}
\begin{center}
\includegraphics[width=8cm]{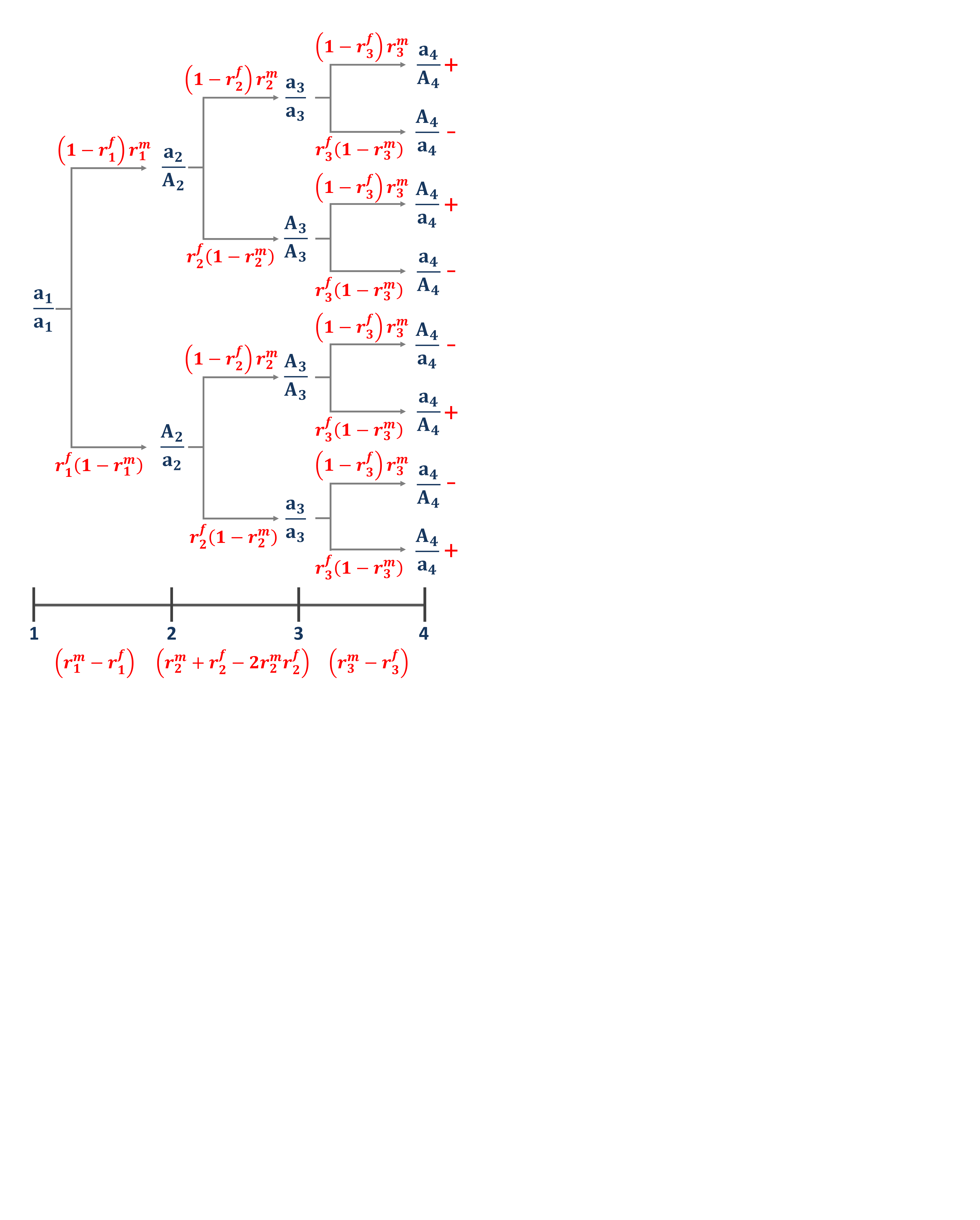}
\caption{{\bf Four-locus tree for the class $HhHh$ of F2 genotypes in the case of sex-specific recombination rates.}}
\end{center}
\end{figure*}

\section{Computer programs for computing probabilities of RIL genotypes}
\label{SM_computer_programs}

Because the $A_{n_1,n_2, \ldots n_L}$ coefficients of the SD equations follow such stereotyped patterns, it is possible to produce a computer program which determines them automatically. We have done so numerically within a C-language code that furthermore uses them to calculate all averages of $k$-spin products recursively for increasing $k$. Once all these averages have been tabulated, the program uses Glauber's equation to compute the probabilities of all $2^L$ RIL genotypes. Note that this last step naively takes on the order of $4^L$ operations, but in fact it is possible to use a multi-dimensional transform that requires only on the order of $L 2^L$ operations~\cite{Zanini_Neher_2012}. The resulting computer program is available online as a Supplementary file. Its computation time grows by about a factor 10 when $L \to L+2$. For illustration, the treatment of the case with $L=14$ loci can be done in less than a second using a standard desktop computer.

To the extent that a purely numerical estimate of RIL probabilities is sufficient, other approaches are also possible. The most straightforward one consists in simulating the steps of production of a RIL, implementing the successive generations until the genotype produced is homozygous. If one repeats this process many times, one can get a large sample of RIL genotypes from which genotype probabilities can be estimated. However the number of different genotypes grows as $2^L$; if one wants to have reliable estimates of all genotypes, the sample must have several hundred realizations of each genotype, no matter how rare each genotype might be. Consequently, this approach is not very useful when the number of loci is greater than 10 and furthermore it is extremely inefficient if one needs precise estimates of the RIL probabilities. To overcome the statistical limitations of stochastic simulation, one may use instead the master equation. The procedure consists in following recursively (from one generation to the next) the probability of all $4^L$ possible genotypes. This recursion can be written as a $4^L \times 4^L$ matrix operating on a vector, and is in fact the approach provided by Haldane and Waddington. The limit of a large number of generations is associated with one of the leading eigenvectors. For $L$ not too large, the appropriate eigenvector can be computed by diagonalizing the matrix, but for large matrices (already at $L=10$ the matrix has more than one million rows), one is forced to rely on the power method. Implementing this method requires one to apply the matrix to the vector a large number of times, large enough to see convergence of the recursion to a fixed point. We have coded this procedure in a C-language program that is also provided online as a Supplementary file. Its computation time grows by about a factor 100 when $L \to L+2$. The treatment of $L=10$ can be done in a few hours on a desktop computer depending on the number of iterations used to get close to the fixed point. It is thus about a million times slower at $L=14$ than the program exploiting the Schwinger-Dyson equations and Glauber's formula.

\end{document}